\documentclass[journal]{IEEEtran}

\usepackage{graphicx}
\usepackage{amsmath}
\usepackage{amssymb}
\usepackage{url}
\usepackage{booktabs}
\usepackage{makecell}
\usepackage{multirow}
\usepackage{cite}
\usepackage{xcolor}

\begin{document}

\title{An Underwater Dehazing Network with Implicit Transmission Estimation}

\author{Sahana~Ray and Sanjay~Ghosh,~\IEEEmembership{Senior Member, IEEE}%
\thanks{S.~Ray and S.~Ghosh are with the Department of Electrical Engineering, Indian Institute of Technology Kharagpur, Kharagpur, WB~721302, India.}%
}
\markboth{Under Review}{Ray and Ghosh: An Underwater Dehazing Network with Implicit Transmission Estimation}
\maketitle

\begin{abstract}
Underwater images suffer from wavelength-dependent light absorption and scattering, which reduces visual quality. This phenomenon could limit the operational reliability of autonomous underwater vehicles, marine surveys, and offshore inspection systems. Purely classical methods often achieve suboptimal performance in real-world datasets, while purely data-driven methods lack physical interpretability. In this letter, we propose UDehaze-iT, a deep network for underwater image enhancement that estimates scene depth implicitly and derives per-channel transmission through the Beer--Lambert law with learnable attenuation coefficients.
We estimate atmospheric light as a semi-classical per-channel scalar, and a zero-initialized residual refiner corrects remaining artefacts after dehazing. To effectively train our method, we apply a composite loss function consisting of five key terms: a $L_1$ loss, a multi-scale patchwise DCT loss, a forward model reconstruction loss, and two regularization terms. With $\sim$0.9M parameters, UDehaze-iT achieves competitive performance on UIEB and UFO-120 datasets. 
\end{abstract}


\IEEEpeerreviewmaketitle

\section{Introduction}
\label{sec:intro}

\IEEEPARstart{L}{ight} propagation underwater differs fundamentally from atmospheric conditions due to the optical properties of water~\cite{duntley1963light,schechner2006recovery}. Long-wavelength components attenuate faster than shorter wavelengths, producing pronounced color casts and low contrast. Scattering from suspended particles introduces haze-like blur that further reduces visibility. Despite this, underwater imaging has become an important resource for ecological research, marine conservation, autonomous underwater vehicles (AUVs) and offshore inspection.

Dehazing and color correction have emerged as the two central problems in underwater image restoration. He~\emph{et al.}~\cite{he2010single} introduced the dark channel prior (DCP) for atmospheric haze, which fails underwater due to severe red-channel attenuation throughout the scene. Akkaynak and Treibitz~\cite{akkaynak2018revised} proposed a revised formation model with explicit wavelength dependence. Peng and Cosman~\cite{peng2015single,peng2017underwater} incorporated blurriness-based depth estimation, and color-correction approaches such as PCDE~\cite{zhang2023underwater} address color casts through segment-wise correction. Retinex-based methods~\cite{land1971lightness,ghosh2019fast,zhang2017underwater,galdran2018duality} decompose the image into illumination and reflectance, while color channel matching (CCM)~\cite{chen2025underwater} normalizes brightness before separately optimizing HSI channels. Among deep-learning methods, Water-Net~\cite{li2019underwater} introduced an encoder-decoder with skip connections for end-to-end enhancement. SCNet~\cite{fu2022underwater} applies spatial and channel normalization in a U-Net backbone. Espinosa~\emph{et al.}~\cite{espinosa2023efficient} combined discrete wavelet skip connections with convolutional block attention.

In this letter, we propose UDehaze-iT, a deep dehazing model that estimates scene depth, attenuation coefficients and atmospheric light to calculate scene radiance, which is then refined further. The primary contributions of this work are:

\begin{itemize}
  \item An encoder-decoder network named \textit{DepthNet} with a dilated bottleneck that estimates a relative depth map without direct depth-supervision; the depth map is converted into a three-channel transmission map $t$ via the Beer--Lambert law with learnable attenuation coefficients.
  \item A convolutional network named \textit{ANet} that learns a residual correction over a classical multi-component prior (combining percentile, dark-channel, and low-variance-patch estimates) to estimate atmospheric or ambient light.
  \item A shallow residual network, named \textit{RefinerNet}, through which the calculated scene radiance is passed; its final layer is zero-initialized, ensuring that early optimization is driven by the dehazing modules.
  \item A \textit{multi-component loss function} that utilizes a multi-scale patchwise DCT loss to preserve depth-correlated frequency content.
\end{itemize}

The remainder of this letter is organized as follows. Section~\ref{sec:method} details the proposed framework, section~\ref{sec:exp} presents the dataset, implementation details, and results, while section~\ref{sec:conclusion} concludes the letter.

\section{Proposed Method}
\label{sec:method}

\subsection{Overview}
\label{ssec:overview}

\begin{figure*}[!h]
\centering
\includegraphics[width=0.9\textwidth]{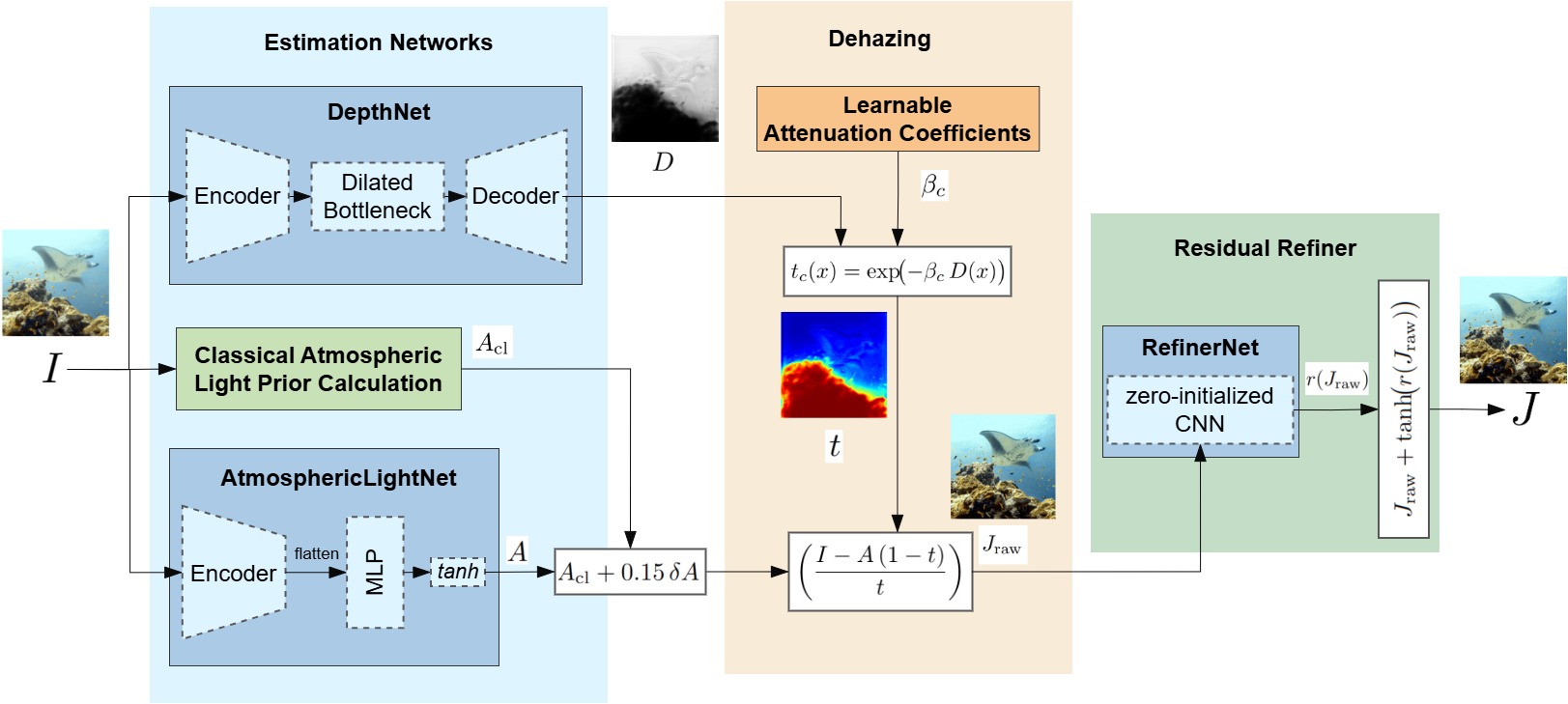}
\caption{Architecture of the proposed model. Given an input image $I$, DepthNet estimates a relative depth map $D(x)$, which is converted into a per-channel transmission map $t$ using learned attenuation coefficients. AtmosphericLightNet predicts the atmospheric light $A$ as a residual correction over a classical prior. A physics-based inversion produces an initial radiance estimate $J_{\mathrm{raw}}$, which is subsequently refined by RefinerNet to yield the final enhanced image $J$.}
\label{fig:arch}
\end{figure*}

We consider the following image formation model:

\begin{equation}
I(x) = J(x)\cdot t(x) + A_c\bigl(1 - t(x)\bigr),
\label{eq:formation}
\end{equation}
where $I(x)$ is the observed degraded image, $J(x)$ is the scene radiance, $t(x) \in (0,1]$ is the transmission map, and $A_c$ is the per-channel atmospheric light. In our proposed method, and estimated depth map $D(x)$ is converted to a three-channel transmission map $t$ via the Beer--Lambert law~\cite{swinehart1962beer} with learned attenuation coefficients.
The initial scene radiance estimate is:
\begin{equation}
  J_{\mathrm{raw}} = \mathrm{clamp}_{[0,1]}\!\left(\frac{I - A_c(1-t)}{t}\right),
  \label{eq:inversion}
\end{equation}
where atmospheric light $A_c$ ($c \in \{R,G,B\}$) is estimated by a dedicated module ANet. Here, $\mathrm{clamp}_{[0,1]}(\cdot)$ denotes element-wise clamping to the range $[0,1]$, i.e., values below $0$ are set to $0$ and values above $1$ are set to $1$, which prevents out-of-range values from propagating. $J_{raw}$ is corrected by a dedicated network RefinerNet to produce the final enhanced output $J$. The full model of the proposed method is illustrated in Fig.~\ref{fig:arch}.

\subsection{DepthNet and Beer--Lambert Transmission}
\label{ssec:depth}

Transmission maps must reflect wavelength-dependent attenuation with depth, yet ground-truth transmission labels are unavailable for real underwater scenes. We therefore introduce \textit{DepthNet}, an encoder-decoder with a dilated bottleneck (dilation rates $d_1$, $d_2$, $d_1$ with $d_1=2$, $d_2=4$) that implicitly estimates a relative depth map, $D(x) \in [d_{\min}, d_{\max}]$, where the bounds $d_{\min}$ and $d_{\max}$ are empirically set to $0.1$ and $10$, respectively. The encoder comprises three stages of strided convolutions, while the decoder restores spatial resolution via bilinear upsampling with skip connections from the corresponding encoder stages. The dilated bottleneck lets DepthNet capture a large receptive field at once---necessary for estimating physically meaningful depth---without losing fine spatial detail.
Per-channel transmission follows the Beer--Lambert law:
\begin{equation}
  t_c(x) = \exp\!\bigl(-\beta_c\, D(x)\bigr), \quad c \in \{R, G, B\}.
  \label{eq:beerlambert}
\end{equation}
where attenuation coefficients $\{\beta_c, \beta_G, \beta_B\}$ are learnable scalars. This exponential decay models the fraction of light at wavelength channel $c$ that propagates through depth $D(x)$.

\subsection{ANet (Atmospheric Light Network)}
\label{ssec:anet}

Classical atmospheric light estimators generalize poorly across scenes with varying turbidity and lighting, yet a purely learned estimator lacks physical grounding. To address this, we design \textit{ANet}, which predicts atmospheric light $A_c$ as a residual correction over a classical prior $A_{\mathrm{cl},c}$ as follows:
\begin{equation}
  A_{\mathrm{cl},c} = \alpha_1\,A_{\mathrm{perc},c} + \alpha_2\,A_{\mathrm{dcp},c} + \alpha_3\,A_{\mathrm{blur},c},
  \label{eq:Acl}
\end{equation}
where $A_{\mathrm{perc},c}$ averages the top $0.1\%$ of brightest pixels in channel $c$, $A_{\mathrm{dcp},c}$ is obtained by selecting the same top $p_{\mathrm{perc}}$ percent of pixels ranked by their dark channel response~\cite{he2010single} (computed over a $15\times15$ neighborhood), and $A_{\mathrm{blur},c}$ is the mean color of the lowest-variance $32 \times 32$ patch in channel $c$. Weights $(\alpha_1, \alpha_2, \alpha_3)$ are empirically set to $(0.5,\, 0.3,\, 0.2)$.

A convolutional network predicts a residual correction $\delta_{A,c} \in (-1, 1)^3$, and the final per-channel estimate is:
\begin{equation}
  A_c = \mathrm{clamp}_{[A_{\min},\,A_{\max}]}\!\bigl(A_{\mathrm{cl},c} + \gamma\,\delta_{A,c}\bigr),
  \label{eq:Anet}
\end{equation}
where $\gamma$ is a scaling factor that limits the network's deviation from the classical prior, and $A_{\min}$, $A_{\max}$ are lower and upper clamping bounds. These parameters are empirically set to $\gamma = 0.15$, $A_{\min} = 0.3$, and $A_{\max} = 1.0$. The scale factor $\gamma$ prevents the learned residual from dominating the physically motivated prior. The lower bound $A_{\min}$ reflects the physical reality that forward-scattered ambient light prevents background pixels from being completely dark underwater.

\subsection{RefinerNet}
\label{ssec:refiner}

Physics-based inversion produces artifacts from imperfect estimates of $t$ and $A_c$. Hence, we introduce a shallow residual network $r(\cdot)$ that corrects artifacts in $J_{\mathrm{raw}}$:
\begin{equation}
  J = \mathrm{clamp}_{[0,1]}\!\bigl(J_{\mathrm{raw}} + \tanh\!\bigl(r(J_{\mathrm{raw}})\bigr)\bigr),
  \label{eq:refiner}
\end{equation}
where the $\tanh$ nonlinearity bounds the additive correction to $(-1, 1)$, preventing the refiner from producing unbounded outputs that would overwhelm the physics-based estimate $J_{\mathrm{raw}}$. The final layer of $r(\cdot)$ is zero-initialized, ensuring that at the start of training $J \equiv J_{\mathrm{raw}}$, and all gradients flow exclusively through the dehazing modules. The refiner therefore only begins making meaningful corrections once $t$ and $A_c$ have stabilized into physically plausible estimates.

\subsection{Loss Function}
\label{ssec:loss}

The total training objective is as follows:
\begin{equation}
  \mathcal{L} = \lambda_1\,\mathcal{L}_{L_1} + \lambda_2\,\mathcal{L}_{\mathrm{DCT}} + \lambda_3\,\mathcal{L}_{\mathrm{fwd}} + \lambda_4\,\mathcal{L}_A + \lambda_5\,\mathcal{L}_\beta,
  \label{eq:total}
\end{equation}
where the loss weights $(\lambda_1, \lambda_2, \lambda_3, \lambda_4, \lambda_5)$ are empirically set to $(1.5,\, 0.8,\, 0.5,\, 0.1,\, 0.1)$. The dominant weight on $\mathcal{L}_{L_1}$ reflects the primacy of pixel-fidelity, while the smaller weights on $\mathcal{L}_A$ and $\mathcal{L}_\beta$ reflect their role as regularizers on intermediate physical quantities rather than direct supervisors of the output.

\subsubsection{$L_1$ Reconstruction}
The output must maintain pixel-level fidelity to the ground-truth reference. We thus supervise training with an $L_1$ reconstruction loss given by:
\begin{equation}
  \mathcal{L}_{L_1} = \|J - J_{\mathrm{GT}}\|_1.
  \label{eq:l1}
\end{equation}
This term enforces pixel-wise fidelity between the enhanced output $J$ and the ground-truth reference $J_{\mathrm{GT}}$.

\subsubsection{Multi-Scale Patchwise DCT Loss}
In underwater images, depth and frequency content are strongly correlated: in shallow regions, the high-frequency signal components survive scattering, whereas in deeper regions, water attenuates them, leaving a region dominated by low-frequency content. We penalize reconstruction error in the DCT domain directly to target this depth-dependent spectral distortion as follows:
\begin{equation}
  \mathcal{L}_{\mathrm{DCT}} = \frac{1}{|\mathcal{P}|}\sum_{P \in \mathcal{P}} w_P\,\frac{1}{3} \sum_{c=1}^{3} \bigl\|\mathrm{DCT}_P(J_c) - \mathrm{DCT}_P(J_{\mathrm{GT},c})\bigr\|_1,
  \label{eq:dct}
\end{equation}
where $w_P \in \{w_s, w_m, w_l\}$ for patch sizes $|\mathcal{P}| \in \{8, 16, 32\}$ pixels, respectively. The patch weights are empirically set to $w_s = 0.1$, $w_m = 0.8$, $w_l = 1.0$, prioritizing larger patches that capture the low-frequency signal structure over smaller ones.

\subsubsection{Atmospheric Light Regularization}
To ensure ANet does not yield  physically implausible values of $A_c$, we introduce:
\begin{equation}
\begin{aligned}
\mathcal{L}_{A} =\ &\|A_c - A_{\mathrm{cl},c}\|_1 \\
&+ \mu_1\,\mathbb{E}\bigl[\mathrm{ReLU}(A_R - A_G + \epsilon) + \mathrm{ReLU}(A_G - A_B + \epsilon)\bigr] \\
&+ \mu_2\,\mathbb{E}\bigl[\mathrm{ReLU}(\bar{I}_c - A_c)\bigr],
\end{aligned}
\label{eq:La}
\end{equation}
where $\bar{I}_c$ denotes the spatial mean of $I$ in channel $c$, $\epsilon = 0.01$ is a small margin, and $(\mu_1, \mu_2)$ are empirically set to $(0.2,\, 0.1)$. Physically, the first term keeps the learned atmospheric light $A_c$ anchored near the classical prior $A_{\mathrm{cl},c}$, preventing the network from exploiting $A_c$ as a free variable that absorbs fitting error. The second term enforces the spectral ordering $A_R \leq A_G$ and  $A_G\leq A_B$, because shorter wavelengths scatter more strongly and thus contribute more to background illumination. The third term penalizes configurations where $A_c$ falls below the mean image brightness $\bar{I}_c$: since the atmospheric light is the dominant illumination source, it should not be less bright than the average of the scene it illuminates. The $\mathrm{ReLU}$ formulation makes each constraint a soft penalty rather than a hard constraint, allowing small violations that the other loss terms may implicitly require.

\subsubsection{Forward Model Reconstruction}
Due to absence of depth or transmission ground-truths, we supervise transmission map with a forward model reconstruction loss as follows:
\begin{equation}
  \mathcal{L}_{\mathrm{fwd}} = \|\hat{I} - I\|_1, \quad \hat{I} = J \cdot t + A_c\,(1-t).
  \label{eq:fwd}
\end{equation}
This term re-applies the underwater image formation model (Eq.~\eqref{eq:formation}) to the estimated clean image $J$, using the predicted transmission $t$ and atmospheric light $A_c$, and penalizes the error between reconstructed observation $\hat{I}$ and input $I$. 

\subsubsection{Attenuation Coefficient Regularization} To ensure estimated attenuation coefficients do not violate the known absorption properties of water, we introduce:
\begin{equation}
\begin{aligned}
\mathcal{L}_\beta =\ &\mathrm{ReLU}(\beta_G - \beta_R + \epsilon) + \mathrm{ReLU}(\beta_B - \beta_G + \epsilon) \\
&+ \nu_1 \sum_c \mathrm{ReLU}(\beta_c - \beta_{\max}) + \nu_2 \sum_c \mathrm{ReLU}(\beta_{\min} - \beta_c),
\end{aligned}
\label{eq:betareg}
\end{equation}
where $\epsilon = 0.01$, and the bounds and coefficients are empirically set to $\beta_{\max} = 2$, $\beta_{\min} = 0.05$, $\nu_1 = \nu_2 = 0.1$. The first two terms enforce the physically required spectral ordering $\beta_R \geq \beta_G$ and $\beta_G \geq \beta_B$: red light is absorbed most strongly per unit depth in water, followed by green and then blue. The third term applies a soft upper bound at $\beta_{\max}$: above this value, the implied attenuation would reduce transmission to near zero over very short distances. The fourth term applies a soft lower bound at $\beta_{\min}$: below this value, the attenuation is negligibly small and the coefficients lose their discriminative physical meaning, potentially collapsing the transmission map.

\begin{figure*}[t]
\centering

\setlength{\tabcolsep}{3pt}
\renewcommand{\arraystretch}{1.1}
\begin{tabular}{cccccc}

\textbf{Input} & \textbf{IBLA}~\cite{peng2017underwater} & \textbf{Shallow-UWnet}~\cite{naik2021shallow} & \textbf{SCNet}~\cite{fu2022underwater} & \textbf{UDehaze-iT} & \textbf{Ground-truth} \\[6pt]

\includegraphics[width=0.15\textwidth]{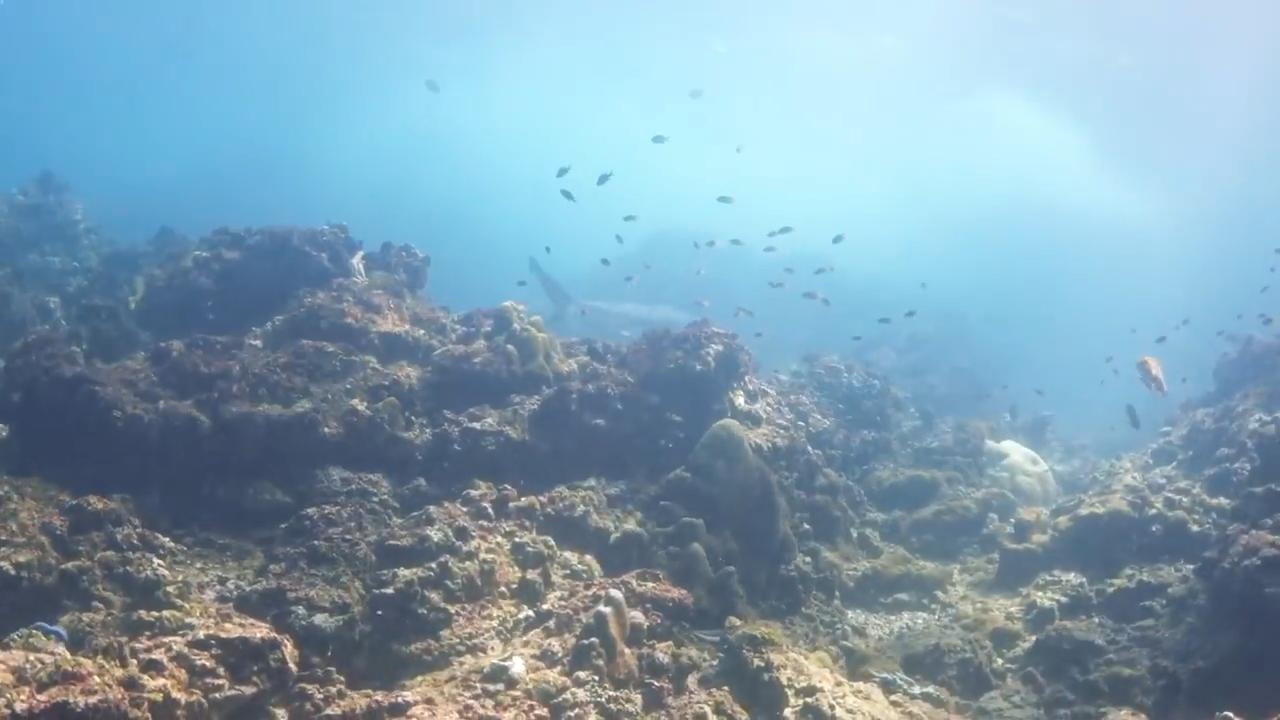} &
\includegraphics[width=0.15\textwidth]{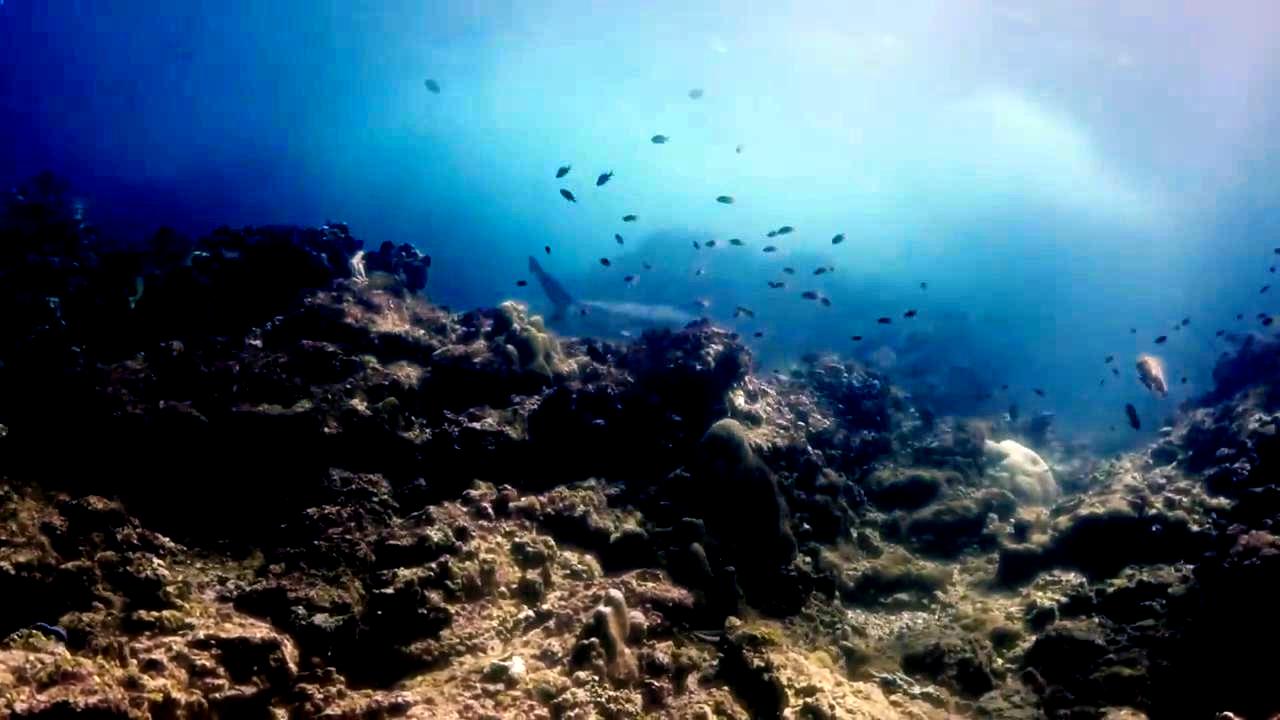} &
\includegraphics[width=0.15\textwidth]{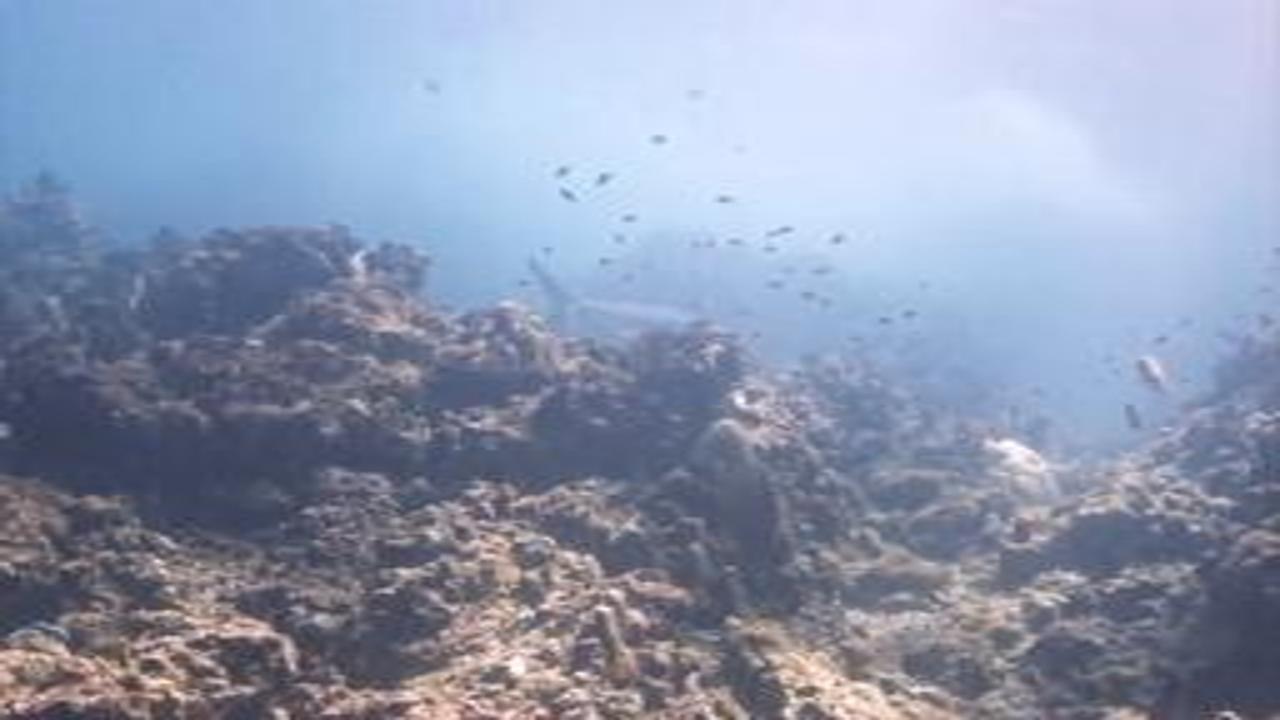} &
\includegraphics[width=0.15\textwidth]{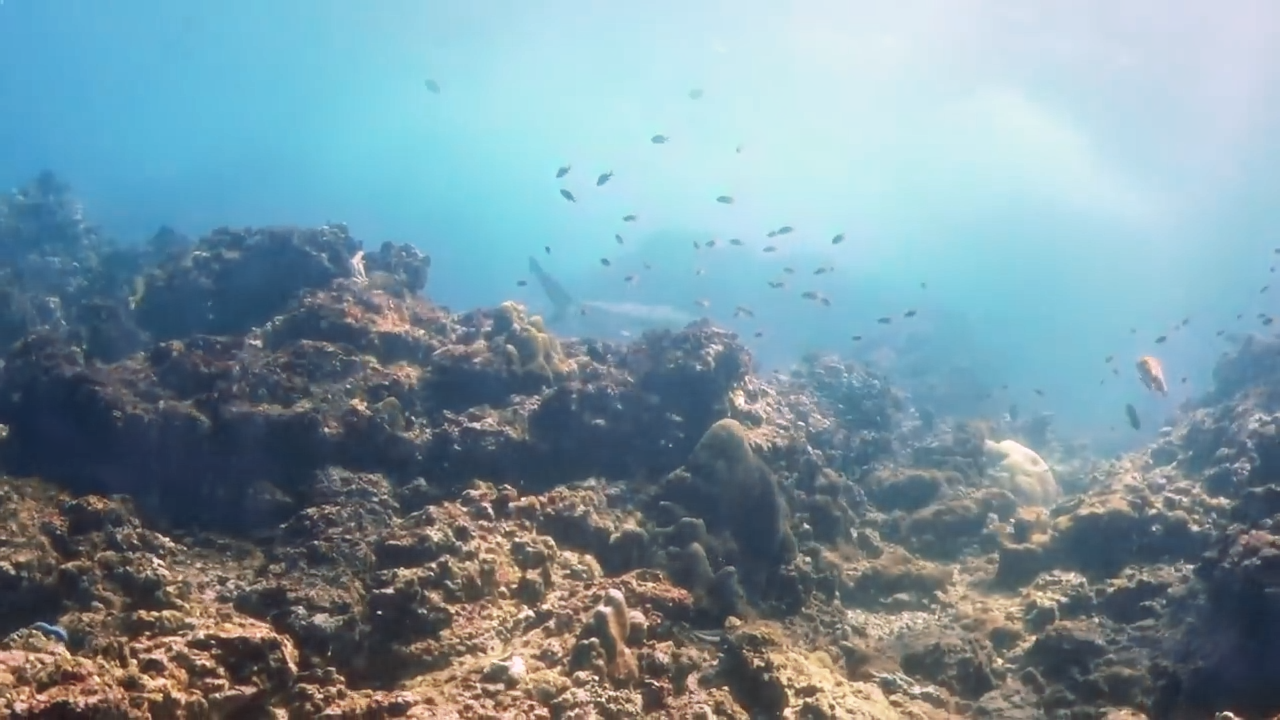} &
\includegraphics[width=0.15\textwidth]{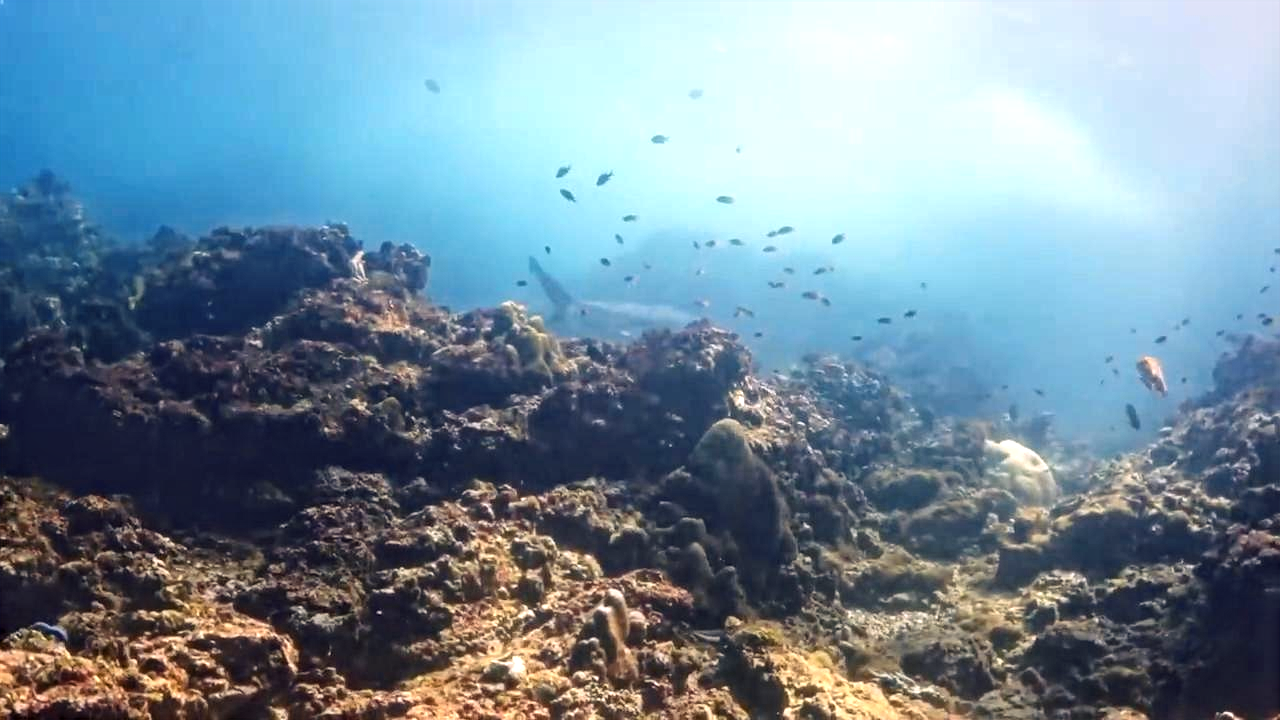} &
\includegraphics[width=0.15\textwidth]{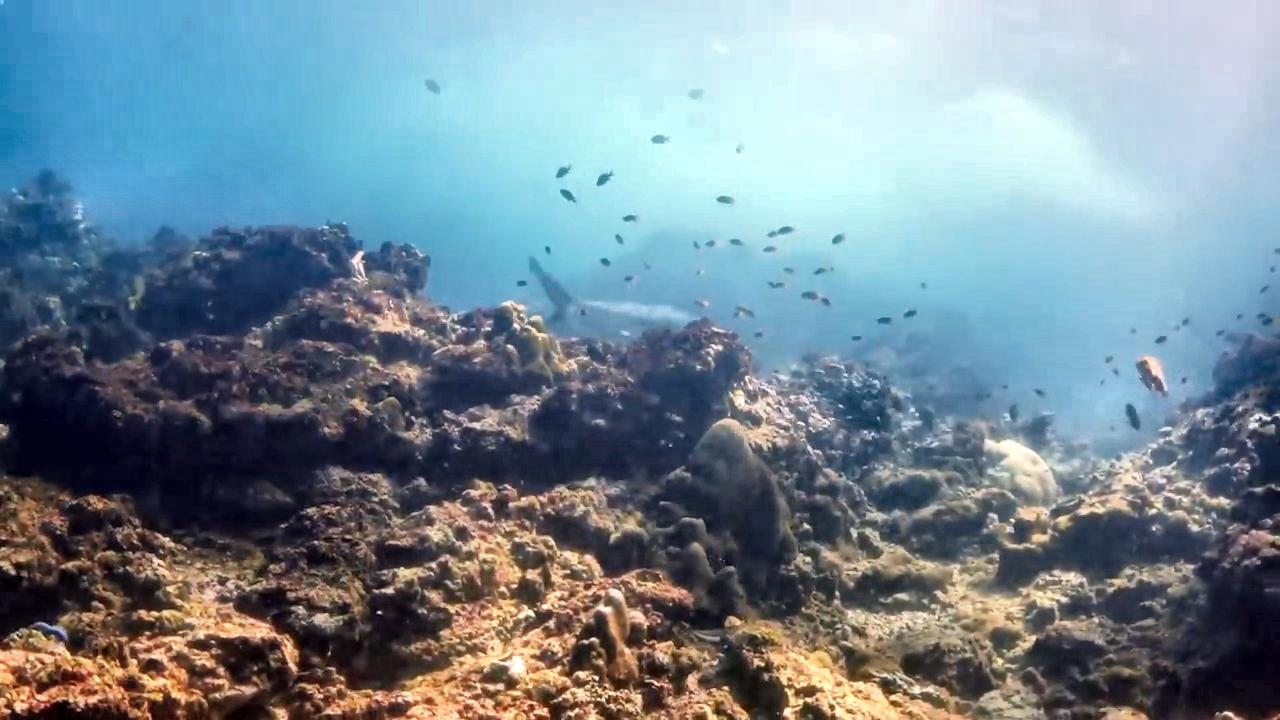} \\[4pt]

\includegraphics[width=0.15\textwidth]{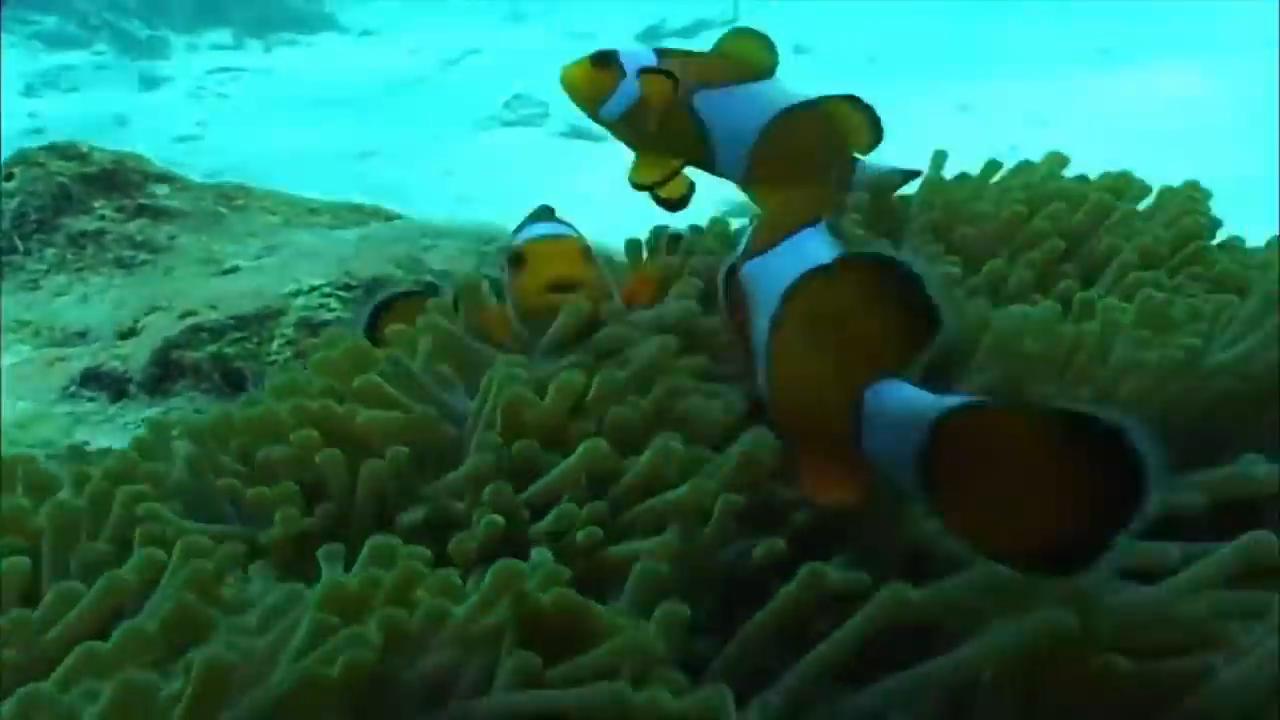} &
\includegraphics[width=0.15\textwidth]{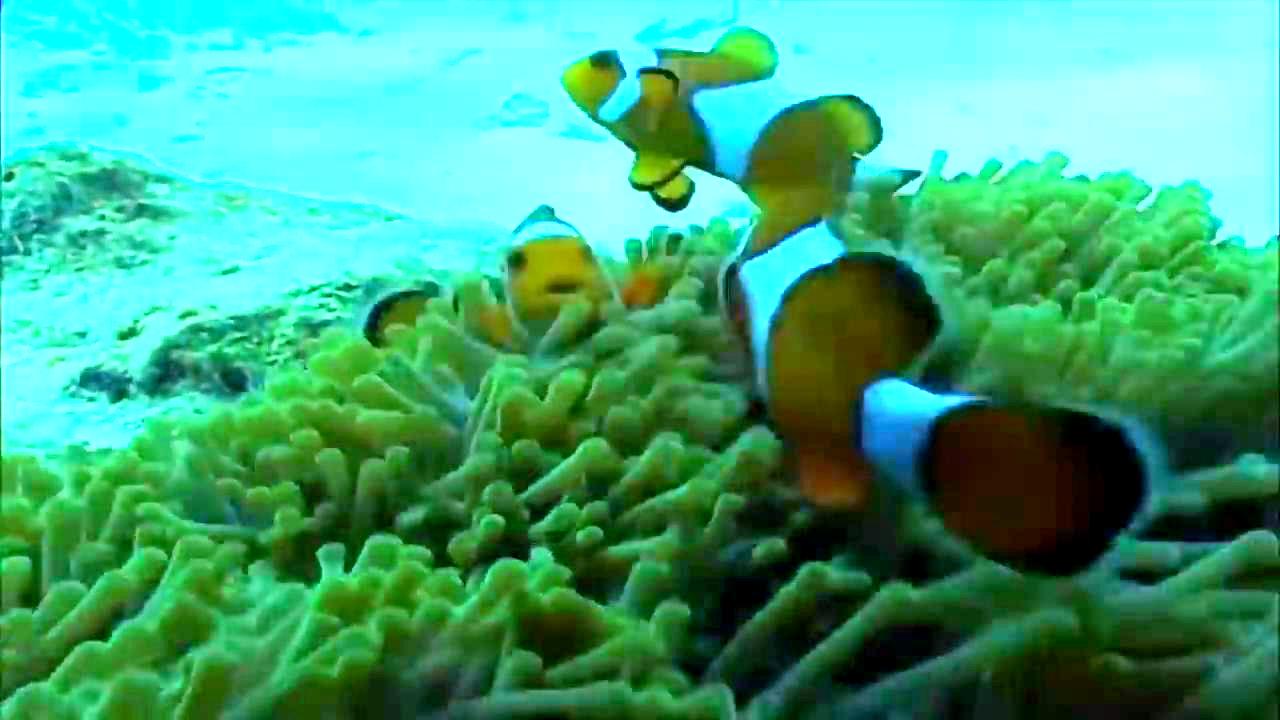} &
\includegraphics[width=0.15\textwidth]{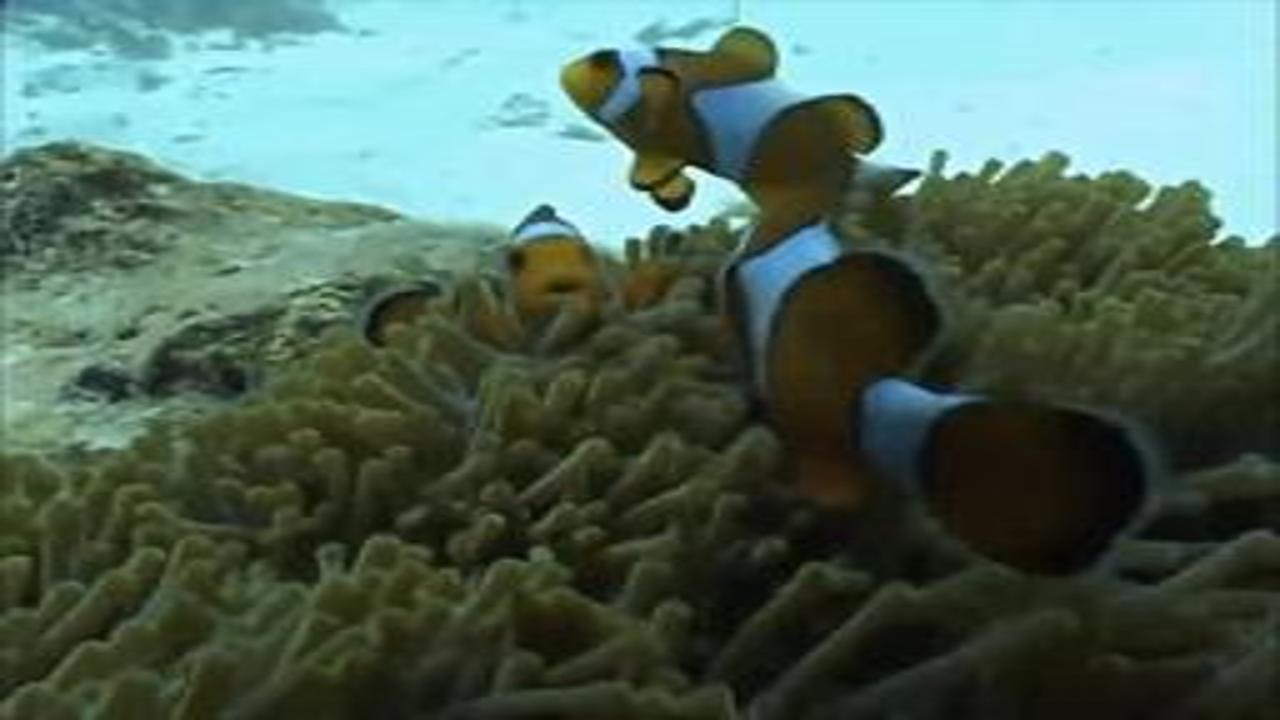} &
\includegraphics[width=0.15\textwidth]{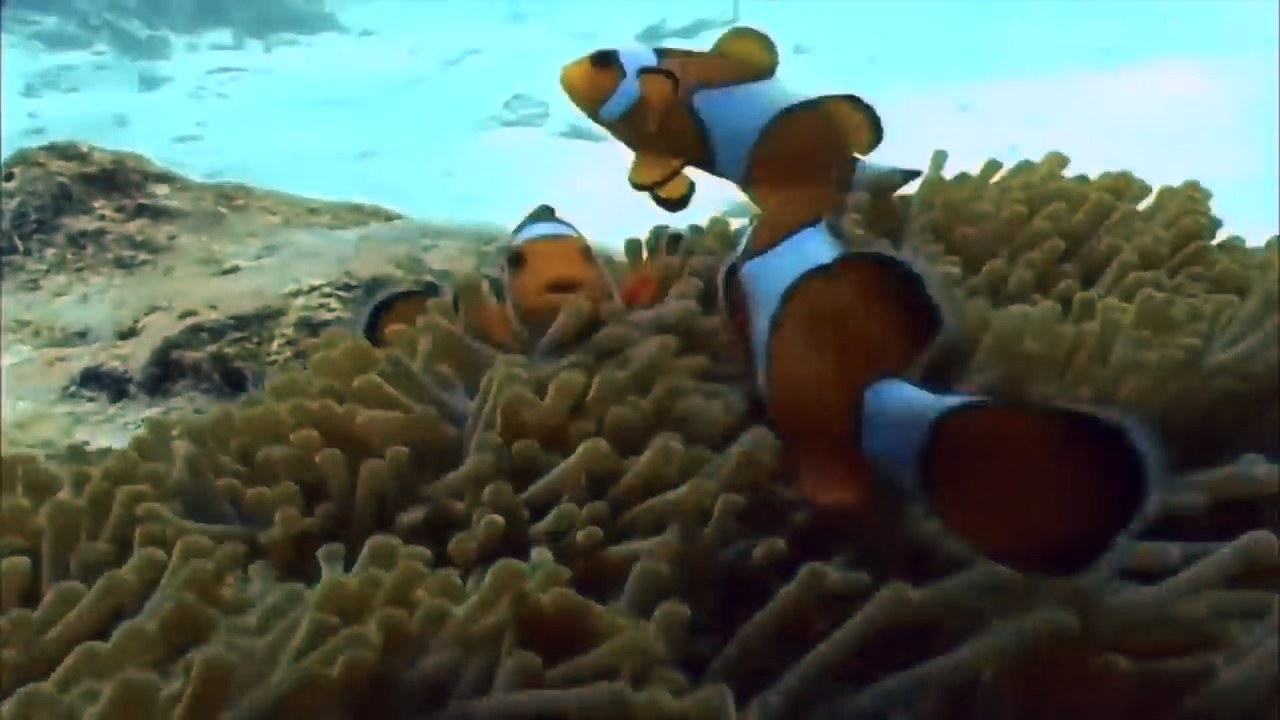} &
\includegraphics[width=0.15\textwidth]{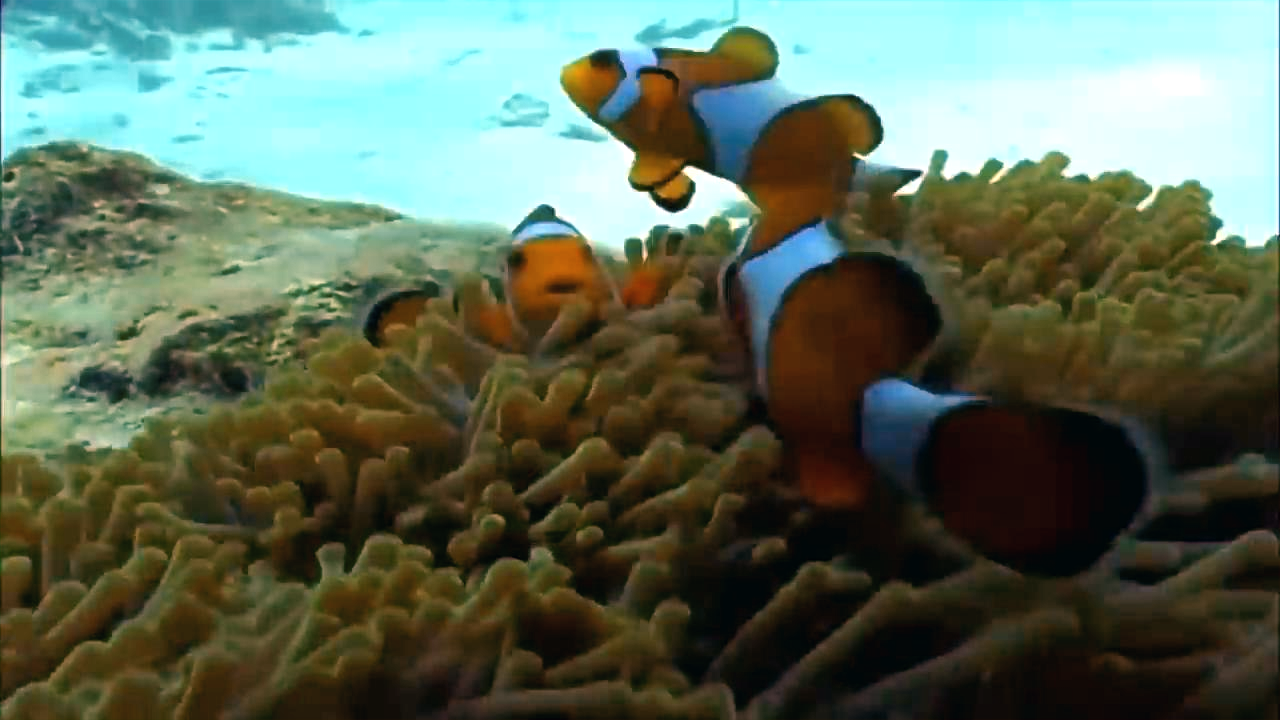} &
\includegraphics[width=0.15\textwidth]{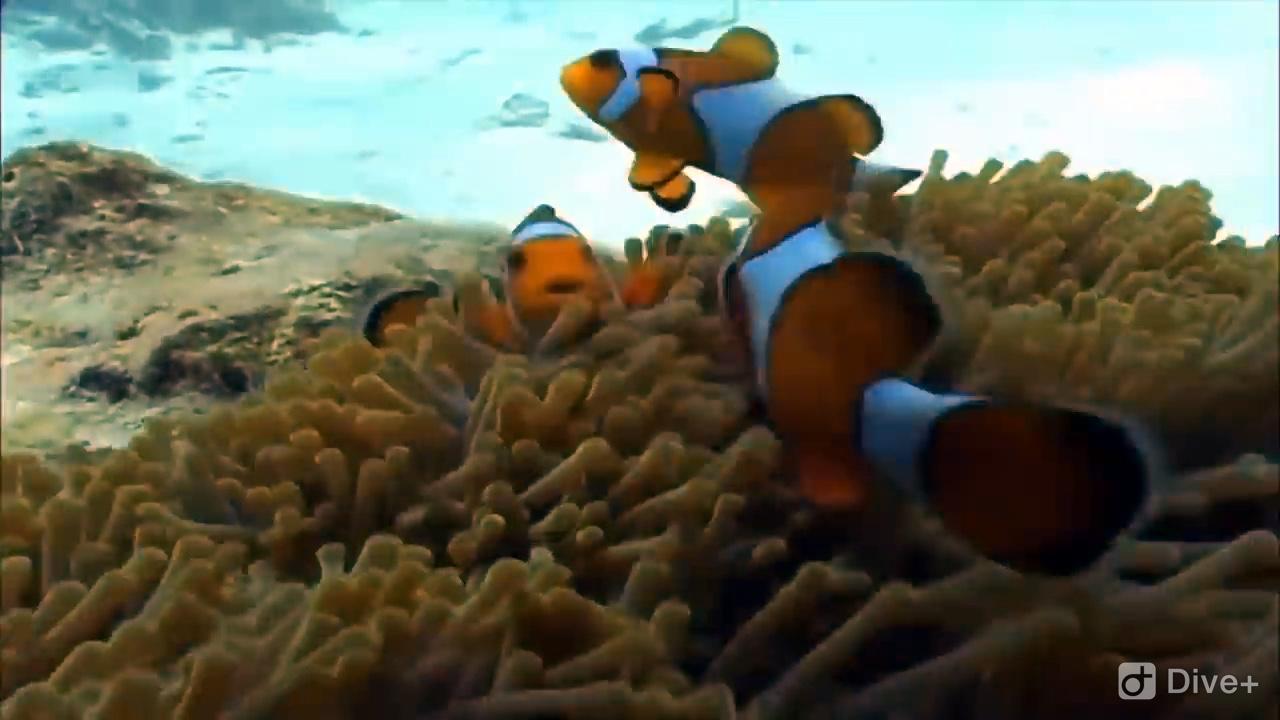} \\[4pt]

\includegraphics[width=0.15\textwidth]{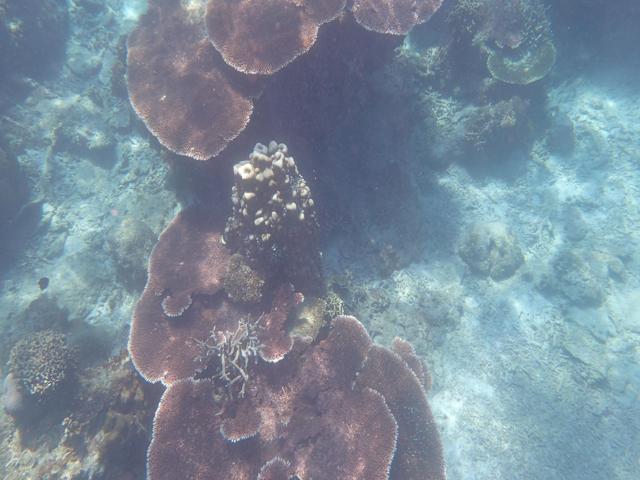} &
\includegraphics[width=0.15\textwidth]{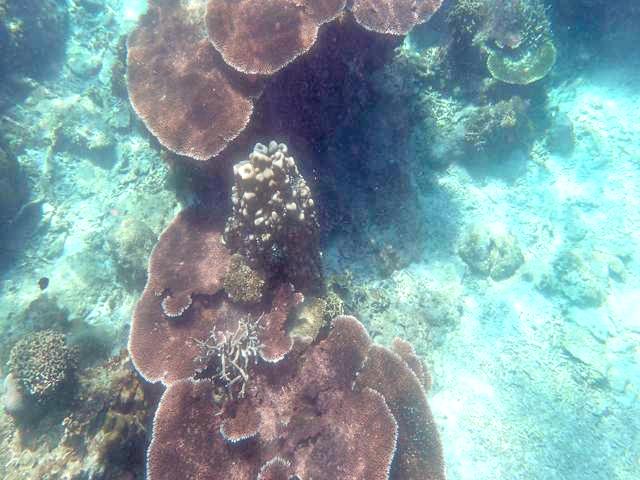} &
\includegraphics[width=0.15\textwidth]{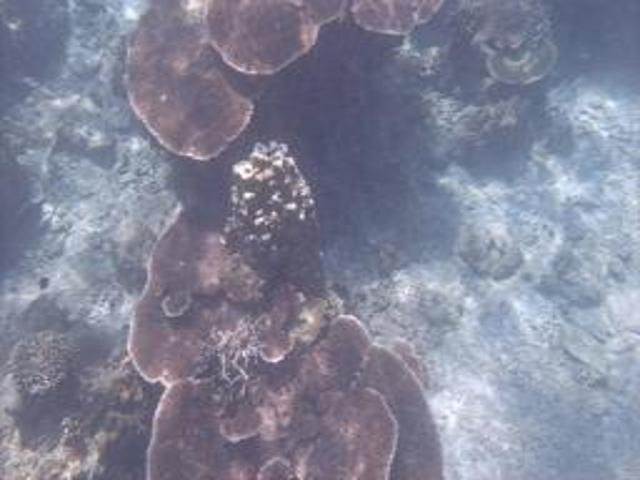} &
\includegraphics[width=0.15\textwidth]{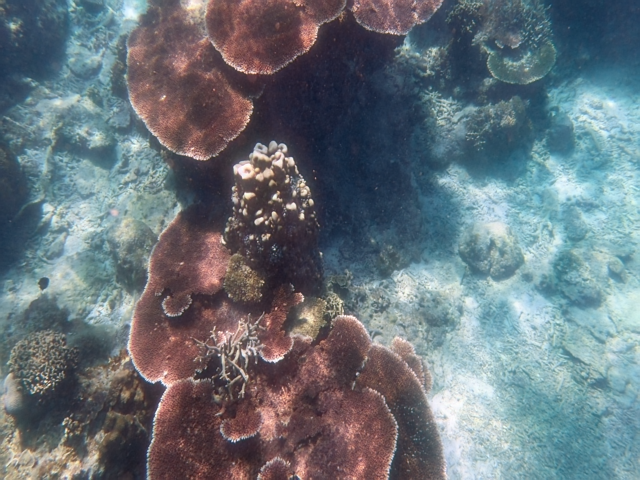} &
\includegraphics[width=0.15\textwidth]{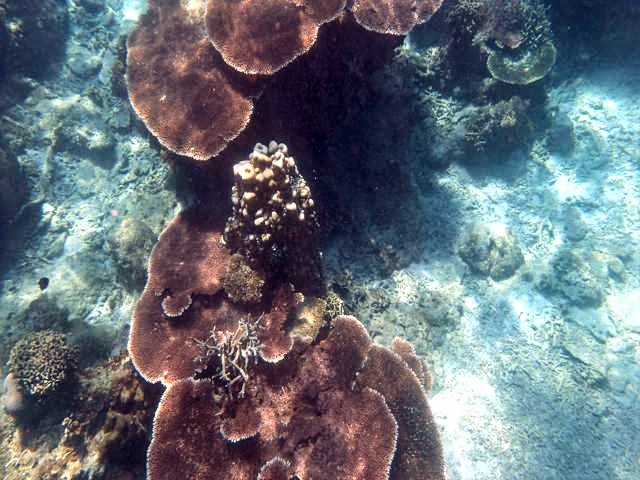} &
\includegraphics[width=0.15\textwidth]{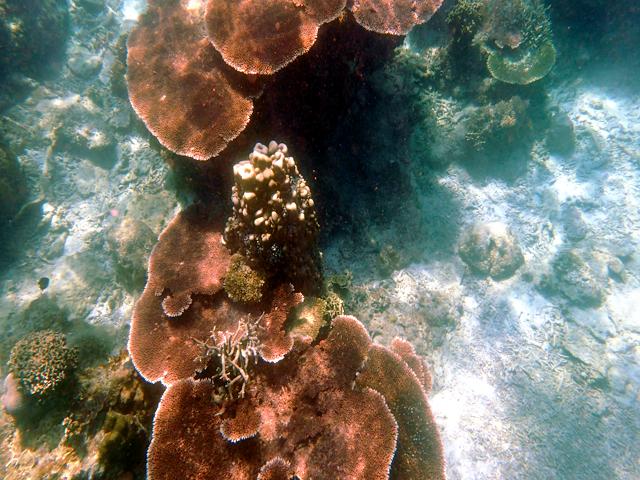} \\[4pt]

\end{tabular}

\caption{Visual comparison of underwater image enhancement results on representative samples from the UIEB dataset. Each row corresponds to a distinct scene, and columns show (left to right) the degraded input, IBLA~\cite{peng2017underwater}, Shallow-UWnet~\cite{naik2021shallow}, SCNet~\cite{fu2022underwater}, the proposed method, and the ground-truth.}
\label{fig:qualitative}
\end{figure*}

\section{Experiments}
\label{sec:exp}

\subsection{Dataset and Implementation Details}
\label{ssec:data}

The proposed model was trained and evaluated on the UIEB~\cite{li2019underwater} and UFO-120~\cite{islam2020fast} datasets.
The UIEB dataset was partitioned into a training subset of the first 700 images and a test subset of 190 images~\cite{fu2022underwater}. All images were resized to $128 \times 128$ pixels to maintain computational efficiency. The AdamW optimizer was used with an initial learning rate of $10^{-3}$ and weight decay $10^{-4}$. The model was trained over 200 epochs with batch size 16 and the best checkpoint was selected by validation PSNR. 


\subsection{Quantitative Results}
\label{ssec:quant}

Model performance was quantitatively assessed using three quality metrics. Peak signal-to-noise ratio (PSNR) measures pixel-level reconstruction accuracy, structural similarity index measure (SSIM)~\cite{wang2004image} evaluates perceptual similarity, and learned perceptual image patch similarity (LPIPS)~\cite{zhang2018unreasonable} measures perceptual dissimilarity in a deep feature space.

\begin{table}[t]
\centering
\caption{Performance comparisons on the UIEB dataset.}
\label{tab:uiebd}
\setlength{\tabcolsep}{10pt}
\begin{tabular}{lccc}
\toprule
Method & SSIM $\uparrow$ & PSNR $\uparrow$ & LPIPS $\downarrow$ \\
\midrule
PCDE~\cite{zhang2023underwater} & 0.7032 & 15.83 & 0.3498\\
IBLA~\cite{peng2017underwater} & 0.5733 & 14.39 & 0.4299 \\
Water-Net~\cite{li2019underwater} & 0.8303 & 19.31 & 0.2016 \\
Ucolor~\cite{li2021underwater} & 0.8412 & 21.97 & 0.1945 \\
Chen~\emph{et al.}~\cite{chen2021underwater} & 0.8770 & 19.37 & 0.1947 \\
Ucolor~\cite{li2021underwater} & 0.8412 & 21.97 & 0.1945 \\
Shallow-UWnet~\cite{naik2021shallow} & 0.8496 & 19.48 & 0.1828 \\
SCNet~\cite{fu2022underwater} & 0.8625 & 22.08 & 0.1936 \\
Espinosa~\emph{et al.}~\cite{espinosa2023efficient} & 0.8802 & 20.92 & -- \\
NU2Net~\cite{guo2023underwater} & 0.8606 & 19.80 & 0.1833 \\
\midrule
\textbf{UDehaze-iT} & \textbf{0.8842} & \textbf{22.10} & \textbf{0.1143} \\
\bottomrule
\end{tabular}
\end{table}

\begin{table}[t]
\centering
\caption{Performance comparisons on the UFO-120 dataset.}
\label{tab:ufo120}
\setlength{\tabcolsep}{10pt}
\begin{tabular}{lccc}
\toprule
Method & SSIM $\uparrow$ & PSNR $\uparrow$ & LPIPS $\downarrow$ \\
\midrule
Water-Net~\cite{li2019underwater} & 0.7330 & 23.12 & 0.3112 \\
Ucolor~\cite{li2021underwater} & 0.7840 & 26.54 & 0.1500 \\
Shallow-UWnet~\cite{naik2021shallow} & 0.8292 & 25.04 & 0.2480 \\
SCNet~\cite{fu2022underwater} & 0.8689 & \textbf{27.12} & 0.2539 \\
NU2Net~\cite{guo2023underwater} & 0.8540 & 25.70 & 0.1612 \\
\midrule
\textbf{UDehaze-iT} & \textbf{0.8724} & 26.62 & \textbf{0.1258} \\
\bottomrule
\end{tabular}
\end{table}

Quantitative comparisons on the UIEB dataset (Table~\ref{tab:uiebd}) demonstrate that the proposed method outperforms all compared methods, achieving 22.10~dB PSNR, 0.8842 SSIM and 0.1143 LPIPS. On UFO-120 (Table~\ref{tab:ufo120}), the proposed method achieves state-of-the-art SSIM and LPIPS, indicating strong cross-dataset generalization, though PSNR is slightly below SCNet. The gains are attributed to two design aspects. First, by modeling depth-dependent attenuation explicitly through the Beer--Lambert law where some parameters are physically constrained. Second, the multi-scale patchwise DCT loss encourages frequency-domain fidelity at multiple spatial scales.

\subsection{Ablation Study}
\label{ssec:ablation}

\begin{table}[!h]
\centering
\caption{Ablation study on loss components on UIEB dataset.}
\label{tab:ablation}
\setlength{\tabcolsep}{10pt}
\begin{tabular}{lccc}
\toprule
Configuration & SSIM $\uparrow$ & PSNR $\uparrow$ & LPIPS $\downarrow$ \\
\midrule
UDehaze-iT (All $\mathcal{L}$) & \textbf{0.8842} & \textbf{22.10} & \textbf{0.1143} \\
w/o $\mathcal{L}_\text{L1}$     & 0.8158 & 18.26 & 0.1692 \\
w/o $\mathcal{L}_\text{DCT}$   & 0.8804 & 21.73 & 0.1218 \\
w/o $\mathcal{L}_{A}$          & 0.8837 & 21.89 & 0.1166 \\
w/o $\mathcal{L}_{\beta}$      & 0.8829 & 21.83 & 0.1184 \\
w/o $\mathcal{L}_\text{fwd}$   & 0.8820 & 21.80 & 0.1189 \\
\bottomrule
\end{tabular}
\end{table}

Table~\ref{tab:ablation} reports the contribution of each loss term by removing one component at a time from the full model. It confirms that $\mathcal{L}_{L_1}$ is the dominant supervision signal. Removing $\mathcal{L}_{\mathrm{DCT}}$ leads to measurable deterioration in both PSNR and LPIPS, consistent with the role of frequency-domain supervision in preserving depth-correlated texture. Although $\mathcal{L}_{\mathrm{fwd}}$ shows moderate effect on metrics, its absence can produce physically inconsistent intermediate representations of $D$ and $t$. The contributions of $\mathcal{L}_A$ and $\mathcal{L}_\beta$ are measurable but lesser, as expected for regularization terms.

\subsection{Qualitative Analysis}
\label{ssec:qual}

Figure~\ref{fig:qualitative} presents representative comparisons on four diverse UIEB test scenes. Across all rows, IBLA consistently over-corrects color, either amplifying existing casts or introducing new ones, and frequently darkens already low-visibility regions. Shallow-UWnet provides modest color correction but produces limited haze removal. SCNet improves overall brightness and color balance, but tends to retain mild residual haze in regions with larger depth variation.

In Rows 1--2 (strong depth variations), our transmission model excels, selectively removing distant haze while enhancing foreground colors for cleaner, more color-balanced results. In Row 3 (coral reef), our method performs well in terms of contrast enhancement and dehazing but shows slightly less color fidelity than other scenes. 

\section{Conclusion}
\label{sec:conclusion}

In this letter, we proposed UDehaze-iT, a deep dehazing network that implicitly estimates relative scene depth and derives per-channel transmission maps via the Beer--Lambert law with learnable attenuation coefficients. Atmospheric light is refined over a classical prior, and a zero-initialized residual refiner corrects dehazing artifacts. Trained with a multi-component loss including frequency-domain supervision and regularization terms, the method achieves state-of-the-art performance on UIEB and UFO-120 datasets. Our method UDehaze-iT could offer a solution for real-world applications such as underwater robotics, and environmental analysis in future.

\bibliographystyle{IEEEtran}
\bibliography{refs}

\end{document}